\documentclass[%
 reprint,
showpacs,%preprintnumbers,
 amsmath,amssymb,
 aps,
 prl,
]{revtex4-1}

\usepackage{graphicx}% Include figure files
\usepackage{dcolumn}% Align table columns on decimal point
\usepackage{bm}% bold math
\begin{document}

\preprint{APS/123-QED}

\title{On the normal modes of weak colloidal gels}% Force line breaks with \\

\author{Zsigmond Varga}

\author{James W. Swan}%
 \email{jswan@mit.edu}
\affiliation{Department of Chemical Engineering, Massachusetts Institute of Technology, Cambridge MA 02139, USA.}%

\date{\today}

\begin{abstract}
The normal modes and relaxation rates of weak colloidal gels are investigated in computations employing different models of the hydrodynamic interactions between colloids. The eigenspectrum is computed for freely draining, Rotne-Prager-Yamakawa and Accelerated Stokesian Dynamics approximations of the hydrodynamic mobility in a normal mode analysis of a harmonic network representing the gel. The spatial structure of the normal modes suggests that measures of collectivity and energy dissipation in the gels are fundamentally altered by long-ranged hydrodynamic interactions, while hydrodynamic lubrication affects only the relaxation rates of short wavelength modes.  Models accounting for long-ranged hydrodynamic interactions exhibit a microscopic relaxation rate for each normal mode, $\lambda$ that scales as  $\lambda\sim l^{-2}$, where $l$ is the spatial correlation length of the mode. For the freely draining approximation, $\lambda\sim l^\gamma$, where $\gamma$ varies between 3 and 2 with increasing $\phi$. A simple phenomenological model of the internal elastic response to normal mode fluctuations is developed, which shows that long-ranged hydrodynamic interactions play a central role in the viscoelasticity of the gel network. Dynamic simulations show that the stress decay as measured by the time-dependent shear modulus matches the normal mode predictions and the phenomenological model. Analogous to the Zimm model in polymer physics, our results indicate that long-ranged hydrodynamic interactions play a crucial role in determining the microscopic dynamics and macroscopic properties of weak colloidal gels.
\end{abstract}

\maketitle

%\tableofcontents

Normal mode analysis (NMA) provides a framework for understanding linear excitations of complex systems. The normal modes are those degrees of freedom which, for small perturbations away from equilibrium, do not interact with each other. Each mode oscillates or relaxes independent of the others with its own characteristic rate. The distribution of normal mode rates, known as density of states (DOS), is central in the investigation of condensed matter, because it allows one to calculate the temperature dependence of the specific heat, thermal conductivity, as well as further thermal and mechanical properties\cite{chen2013phonons}. NMA is also a basis for predicting liquid-state dynamics\cite{seeley1989normal}, biomolecular relaxation\cite{bahar2009normal}, the jamming transition of granular materials\cite{prasad2003rideal}, and the phonon density of colloidal crystals and glasses\cite{keim2004harmonic, ghosh2010density}.  

While significant effort has been devoted to the investigation of the DOS of dense glasses and other disordered colloidal solids\cite{kaya2010normal}, much less is known for the case of colloidal gels. In attractive colloidal dispersions particles may aggregate and kinetic arrest can result. The dispersion follows a kinetic pathway that is dependent on volume fraction, interaction strength, and range of interaction\cite{Zaccarelli2007}. The rheological and structural properties of colloidal gels make them highly desirable for many technological applications\cite{Gaponik2011}. Additionally, the normal modes of soft materials and their associated DOS may be directly related to their linear viscoelastic response\cite{tschoegl}. Developed frameworks aim to connect the spectrum of relaxation times to the macroscopic rheology\cite{zaccone2009elasticity, zaccone2014linking}.

Recent work has used simulations of microstructural dynamics and relaxation to predict mechanical properties of gels\cite{rovigatti2011vibrational, taraskin1997nature, krall1998internal, ilyin2009randomness, russo2009reversible}. Such work neglects the role of hydrodynamic coupling of the suspended particles. Contrary to atomic systems, where the phonon dispersion curve is entirely determined by particle mass and inter-particle potentials, in colloidal systems dissipative interactions mediated by the suspending fluid must be considered \cite{baumgartl2008phonon, royall2015probing}. Because of their long-range nature these interactions are difficult to treat theoretically. The complexity of modeling many-bodied hydrodynamic interactions (HI) in large systems has encouraged their neglect in work to date. However, as was recently shown, neglecting long-ranged dissipative coupling in discrete element modeling of gelling systems leads to predictions of structure and dynamics that are at odds with experimental observations\cite{varga2015hydrodynamics, varga2016hydrodynamic}. Indeed, since the modes of relaxation are modulated by the hydrodynamic interactions between particles, the fluid mechanics within a gel must be central to the linear viscoelastic moduli they exhibit when deformed macroscopically. Several experiments \cite{hurd1982lattice, keim2004harmonic} have found that models neglecting hydrodynamic coupling fail to reproduce measured dynamics and friction coefficients. There is a need for a hydrodynamic theory to explain the observed relaxation modes \cite{baumgartl2008phonon, whitaker2016bond}. Hurd et al. \cite{hurd1985friction} developed a systematic transport theory for normal-modes in a harmonic lattice of colloidal particles immersed in a viscous medium and obtained results for dilute colloidal crystals. For amorphous, space spanning configurations, such as colloidal gels, no analytical solution is available. 

We have recently developed methods for rapid calculation of hydrodynamic interactions in suspensions of mono-disperse spheres\cite{Swan2015, varga2015hydrodynamics, fiore2016rapid}. Varying levels of approximation are possible for any particle configuration and system sizes up to $10^6$ particles. We conduct simulations of colloidal dispersions with $N=10^2$ --$10^4$ particles of radius $a$ for up to $10^4$ bare diffusion steps $\tau_D=6\pi\eta a^3/k_B T$. A short-range attraction, mimicking the polymer induced depletion attraction in experimental systems \cite{russel1989colloidal, Poon1997, Lu2006}, is given by an Asakura-Oosawa form \cite{Asakura1958}. The range of the attraction is set to $0.1a$ and its strength, proportional to the polymer concentration, is $-10k_B T$ at contact. For each of the three volume fractions studied, $\phi=0.15\%, 30\%$ and $45\%$, we generate 5 independent configurations\cite{varga2015hydrodynamics}. At the end of each simulation the sample is gelled, with all particles belonging to a percolating cluster.

We use particle positions from simulations for NMA to determine the distribution of microscopic relaxation rates of the colloidal gel. Evolution of the positions $\mathbf{x}(t)$ is dictated by the over-damped momentum balance: $ {\dot{\mathbf{x}} = -\mathbf{M}  \left(\nabla U - \mathbf{F}_B\right)} $, where $\mathbf{M}$ is the hydrodynamic mobility, $U$ is the inter-particle potential, and $\mathbf{F}_B$ is the Brownian force satisfying the fluctuation-dissipation theorem\cite{Edwards1986}. The kinetically arrested gel state is close to mechanical equilibrium and the dynamics are nearly harmonic. Denoting particle fluctuations about local mechanical equilibrium, $\mathbf{x}_{eq}$, as $\boldsymbol{\delta}=\mathbf{x}-\mathbf{x}_{eq}$, the time evolution of an average fluctuation is to leading order: $\boldsymbol{\dot{\delta}} = -\left(k/(6\pi\eta a)\right) \mathbf{\hat{M}} \mathbf{\hat{H}}\boldsymbol{\delta}$. $\mathbf{\hat{M}}$ is the mobility in the $\mathbf{x}_{eq}$ configuration scaled on the Stokes drag $6\pi\eta a$ and $\mathbf{\hat{H}}$ is the Hessian of $U$ in the $\mathbf{x}_{eq}$ configuration scaled on the characteristic bond stiffness, $k$. This shadow system \cite{henkes2012extracting}, represents the gel as a network of harmonic springs connecting particles in the configuration, $\mathbf{x}_{eq}$. The perturbation may be decomposed into normal modes: $\boldsymbol{\delta}(t) = \sum^{3N}_{i=1} C_i  \mathbf{v}_i e^{-\lambda_i t}$, for which: 
\begin{equation}
\lambda_i \mathbf{\hat{R}}\mathbf{v}_i= \mathbf{\hat{H}}\mathbf{v}_i.
\label{nma}
\end{equation}  
The NMA relates normalized displacements $\mathbf{v}_i$ and velocities $\lambda_i\mathbf{v}_i$ set by the microscopic relaxation rate $\lambda_i$ of mode $i$. $\lambda_i$ is normalized by $6 \pi \eta a/k $ and $\mathbf{\hat{R}}=\hat{\mathbf{M}}^{-1}$ is the hydrodynamic resistance.  The set $\mathbf{v}_{i} $ for $ i=1,...,3N $ is made orthogonal by taking a product with the Cholesky factorization of $\mathbf{\hat{R}}$. 

The resistance tensor is computed using the Accelerated Stokesian Dynamics (ASD) method \cite{sierou2001accelerated, banchio2003accelerated} for any particle configuration. In this framework, $\hat{\mathbf{R}}$ is a superposition of a far-field contribution due to long-ranged HI and a near-field component accounting for pair-wise lubrication forces: $\mathbf{\hat{R}} = \mathbf{\hat{R}}_{FF}+\mathbf{\hat{R}}_{NF}$. However, resistance problems are ill-conditioned and solutions are computationally expensive to evaluate accurately\cite{kim2013microhydrodynamics}. Lubrication is dominated by squeezing flows generated by motion along the axis of pairwise particle separations.  For nearly touching particles, $ \hat{\mathbf{R}}_{NF} $ is approximately congruent to the Hessian of the potential; therefore, we make the following simplifying assumption: $\mathbf{\hat{R}}_{NF} \approx \alpha \mathbf{\hat{H}}$, such that $\mathbf{\hat{R}}=\mathbf{\hat{R}}_{FF}+\alpha\mathbf{\hat{H}}$. The parameter $\alpha$ can be approximated as the relaxation time for a spring connecting nearly touching particles, which when made dimensionless on $ 6 \pi \eta a / k $ is simply:  $a/\xi$, where $\xi=r_{eq}-2a$ is the equilibrium separation between neighboring particle surfaces. If this hypothesis holds, solutions to the renormalized eigenvalue problem: $\lambda_{FF}  \mathbf{\hat{R}}_{FF}\mathbf{v}_i = \mathbf{\hat{H}}\mathbf{v}_i$, have the same eigenvectors as \eqref{nma} and eigenvalues that are related by: $\lambda^{-1}-\lambda_{FF}^{-1}=\alpha$. Consequently, the effect of hydrodynamic lubrication is to reduce the relaxation rates of normal modes, $\lambda_i$, but not to change the spatial structure of the modes, $\mathbf{v}_i$. The characteristic timescale $\alpha$ relates the relaxation spectrum of models neglecting hydrodynamic lubrication to those including it!

\begin{figure}[h!]
\center
\includegraphics[width=\columnwidth]{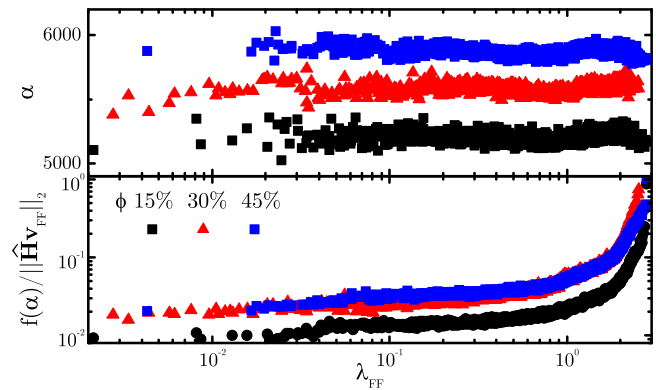}
\caption{Top: $\alpha$ computed by minimizing $ f( \alpha ) $ with $N=500$. Bottom: The corresponding approximation error as a function of $\lambda_{FF}$.  \label{fig:asd}}
\end{figure}

For 500 particles, we compute $\mathbf{\hat{R}}$ from a full ASD simulation and $ \mathbf{\hat R}_{FF} $ using the inverse of the Rotne-Prager-Yamakawa (RPY) mobility, which treats the particles as renormalized Stokeslets, and is the leading order far-field contribution to the hydrodynamic interactions among suspended particles\cite{rotne1969variational}. We obtain $\alpha$ by calculating the value that minimizes:
\begin{equation}
 f(\alpha) = ||\frac{\lambda_{FF}}{1+\alpha\lambda_{FF}}\mathbf{\hat{R}}\mathbf{v}_{FF}-\mathbf{\hat{H}}\mathbf{v}_{FF}||_2,
 \end{equation} 
 for each mode determined via the RPY model $ \{\lambda_{FF}, \mathbf{v}_{FF} \} $, and plot the result as a function of $\lambda_{FF}$ in figure \ref{fig:asd}.  On this figure, we also depict $ f( \alpha ) / || \mathbf{\hat H}  \mathbf{v}_{FF} ||_2 $, a normalized error in approximation of the modes of the ASD simulation by $ \{\lambda_{FF}/(1+\alpha \lambda_{FF}), \mathbf{v}_{FF}\} $.  The computed value of $ \alpha $ is nearly constant for all the modes at a given volume fraction.  As $\phi$ increases, so does $ \alpha $.  This may be interpreted as the effective equilibrium separation between particles, $\xi$, decreasing slightly at higher volume fractions.  Direct computation of the average surface to surface separation yields values consistent with $ \alpha $ at each volume fraction.

These results suggest that for virtually all the modes computed directly via the RPY model, the approximation $\mathbf{\hat{R}}_{NF}=\alpha\mathbf{\hat{H}} $ holds. Slowly relaxing modes involve collective motion of the particle network and are unaffected by localized squeezing flows.  Larger normalized errors are found for modes with $ \lambda_{FF} > 1 $.  As we will show, these fast modes involve localized rather than collective motion.  As a consequence, similarity of eigenvectors in this part of the relaxation spectrum breaks down.  In the supplementary material, we examine the effect of system size and show that the proportion of fast modes and the error in $ f( \alpha ) $ over the spectrum shrinks with increasing $ N $.  Thus, the proposed renormalization may become exact in the limit $ N \rightarrow \infty $.  We now examine how long-ranged HI affect relaxation in colloidal networks. We compare the normal modes obtained from the freely draining approximation (FD), where $\mathbf{\hat{M}}=\mathbf{I}$ with that obtained from the RPY tensor with $ N = 10^4 $ in order to fully describe the distribution of slow modes responsible for macroscopic viscoelasticity via the density of states. Images and structure factors of the gels are in the supplementary material.

\begin{figure}[h!]
\center
\includegraphics[width=\columnwidth]{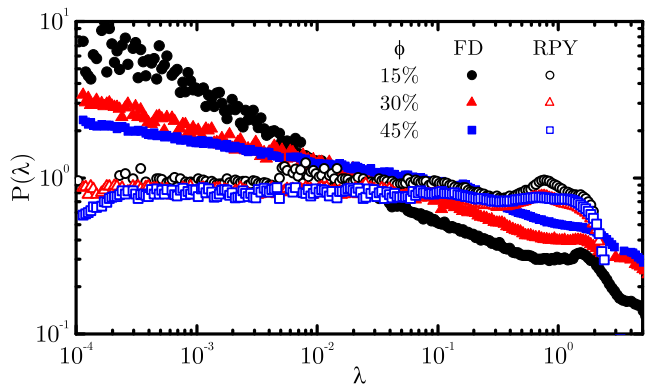}
\caption{Density of states as a function of relaxation rate $\lambda$, $P(\lambda)$ for all gels, modes and models under study. \label{fig:omega}}
\end{figure}

In figure \ref{fig:omega} we plot the DOS obtained for each volume fraction with the FD and RPY models. The DOS as a function of relaxation rate, $P(\lambda)$, highlights the significant effect of long-ranged HI on the normal mode response in weak colloidal gels. For the FD model, we find an overabundance of slow modes. The DOS scales as $P(\lambda)\sim\lambda^{-\beta}$, where $\beta\approx-1/2$ for $\phi=15\%$. In contrast, for the RPY spectrum, the occurrence of slow and fast relaxation modes is equally probable, and $\beta=0$ for all $\phi$. With increasing $\phi$, differences between the two hydrodynamic models diminish and $\beta$ approaches zero as the fraction of slow modes decreases for the FD model. This is analogous to differences observed in polymer physics. A dilute solution of ideal polymer chains obeying the Zimm model, has $P(\lambda)\sim\lambda^{-1/3}$. In contrast, when HI are neglected as in the Rouse model, $P(\lambda)\sim\lambda^{-1/2}$\cite{Edwards1986}. HI favor a flatter distribution of relaxation rates as collective motion accelerates stress relaxation in soft materials.  In the supplementary material we compare the DOS for the same gels using a hydrodynamic model that accounts for higher order multipoles of the force density on the particle surfaces. From this, we conclude that the DOS is largely unaffected by higher order hydrodynamic couples.

\begin{figure}[h!]
\center
\includegraphics[width=\columnwidth]{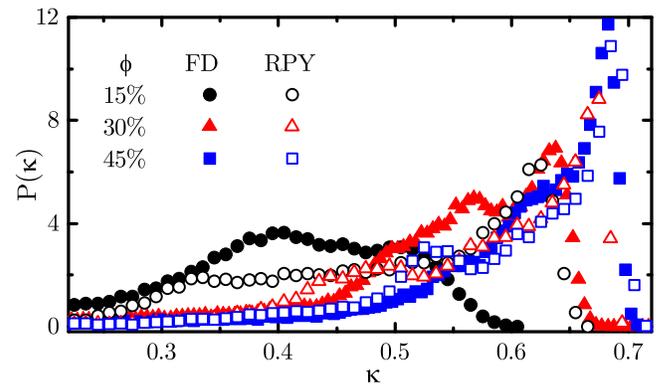}
\caption{Probability density for collectivity index of all gels, modes and models under study. \label{fig:dispersion}}
\end{figure}

To characterize the structure of the modes we compute the collectivity index\cite{bahar2009normal}, 
\begin{equation}
\kappa_i=\frac{1}{N} \exp \left(-\sum^N_{j=1}\mu |\mathbf{v}_i^{(j)}|^2 \log\left(\mu|\mathbf{v}_i^{(j)}|^2\right) \right)
\end{equation}
where $\mu$ is the normalization constant: $ \mu^{-1} = \sum_j |\mathbf{v}_i^{(j)}|^2$, and $ \mathbf{v}_i^{(j)} $ is the part of mode $ i $ corresponding to particle $ j $. The collectivity index measures the degree to which particles participate in each mode. A mode that excites a larger number of particles in the gel, has the higher collectivity index. With increasing volume fraction the degree of collectivity increases due to increased rigidity. In figure \ref{fig:dispersion}, we plot the likelihood of observing a mode with collectivity index $\kappa_i$ for all gels and both hydrodynamic models under study. 
 
We find that the normal modes are more likely to have higher participation in RPY model. Long-ranged HI lead to coupling and excitation of a larger fraction of the particles. As the volume fraction increases, differences between the two hydrodynamic models diminish. For $\phi=0.45$, particle crowding results in hydrodynamic screening, and $P(\kappa_i)$ is similar for both FD and RPY. The small $\kappa_i$ tail of the distributions corresponds to fast modes, $\lambda>1$, for which relaxation is much faster than $6\pi\eta a/k$, and the normal modes are localized to only a few particles. 

\begin{figure}[h!]
\center
\includegraphics[width=\columnwidth]{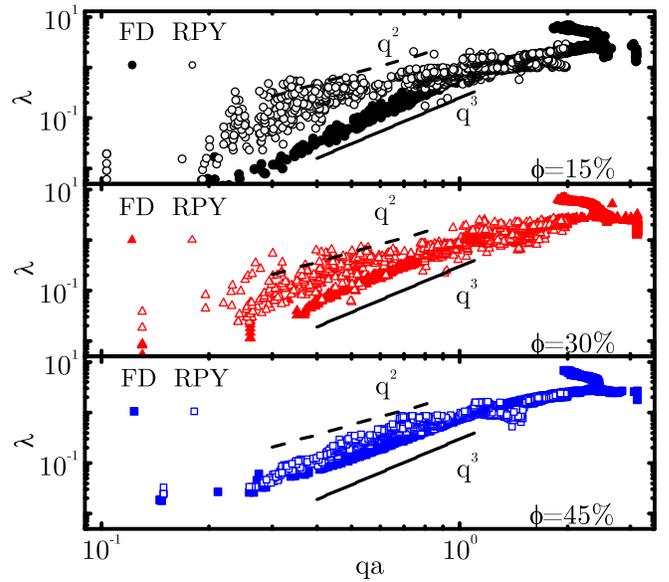}
\caption{Relaxation rate $\lambda$ as a function of the corresponding characteristic wavelength $q$ for all modes, gels and both models. \label{fig:energy}}
\end{figure}

We also compute the correlation length of each normal mode through a quantity used in the analysis of energy densities in turbulent flow\cite{sirovich1987turbulence}: 
\begin{equation}
S_{\mathbf{v}_i}(\mathbf{q})=\lambda_i \sum^N_{j,k=1}\mathbf{v}_i^{(j)}\cdot\mathbf{v}_i^{(k)} \exp\left( i\mathbf{q} \cdot (\mathbf{x}^{(k)}_{eq}-\mathbf{x}^{(j)}_{eq}) \right),
\end{equation}
with $ \mathbf{x}^{(j,k)}_{eq} $ the position of particle $ j $ or $ k $ in the $ \mathbf{x}_{eq} $ configuration.  $S_{\mathbf{v}_i}(\mathbf{q})$ is analogous to the Karhunen-Lo\`{e}ve decomposition, and quantifies the energy dissipated in eigenmode $\mathbf{v}_i$ over wavelength $\mathbf{q}$. The local maximum in $S_{\mathbf{v}_i}(\mathbf{q})$ , denoted $q_i$, determines the characteristic length scale $l_i = 2\pi/q_i$ of energy dissipation in the gel structure with relaxation rate $\lambda_i$. Figure \ref{fig:energy} plots $\lambda_i$ versus $q_i$ for all the gels, models and modes under study. We observe a power-law trend $\lambda\sim q^\gamma \sim l^{-\gamma}$ that depends on the hydrodynamic model. For the RPY model, scaling of relaxation rate with spatial correlation length is independent of $\phi$ and $\gamma\approx 2$. The FD modes exhibit significant dependence on volume fraction. For $\phi=0.15$, we find $\lambda\sim l^{-3}$, while at higher volume fractions, $\gamma$ decreases and approaches the behavior exhibited by the RPY model.

The dispersion relation can be understood as the ratio of elasticity to viscous dissipation. The elastic response of a domain of size $l$ is characterized by a spring constant $k(l) \sim l^{-z}$, where $z$ is the elasticity exponent. With increasing size $l$, a smaller fraction of bonds connect the domain, which leads to softening and a decreasing $k(l)$. $z$ is sensitive to the nature of the bonds and the resulting stiffness of the gel backbone\cite{meakin1984topological}. In our study we consider only central forces between the particles which leads to no bending rigidity and an exponent $z<3$ that will depend solely on the backbone structure at given $\phi$\cite{kantor1984elastic}. The viscous dissipation however will depend on the hydrodynamic model chosen. We therefore expect the normal mode relaxation rate to decay differently with increasing correlation length. For the FD model, the drag on a domain of size $l$ scales as $\sim l^{d_f}$,  so that $\lambda_{FD}\sim l^{-(d_f+z)}$, the relaxation rate is given by the ratio of spring constant to drag coefficient\cite{krall1998internal}. For the RPY model, the drag is linear in $l$ such that $\lambda_{RPY}\sim l/^{-(1+z)}$. From the $\lambda \sim l^{-2}$ scaling of the RPY spectrum we find that the elasticity exponent of the weak gels under study is $z\approx 1$. Employing the box counting method we determine that the fractal dimension of the gels at low $\phi$ is $d_f=2.1$, which when added to the same value of $z$, yields the observed scaling exponent for the FD dispersion relation.

As the volume fraction increases, hydrodynamic screening makes the effective drag coefficients for the two models similar and yields comparable dispersion relations. The fraction of particles with 6 neighbors or higher increases to more than 92\% for $\phi=0.45$, and $z\rightarrow 0$ as the number of free chains leading to softening over larger length scales vanishes in the now rigid structure. Consequently $\lambda\sim l^{-2}$ for both models.

\begin{figure}[h!]
\center
\includegraphics[width=\columnwidth]{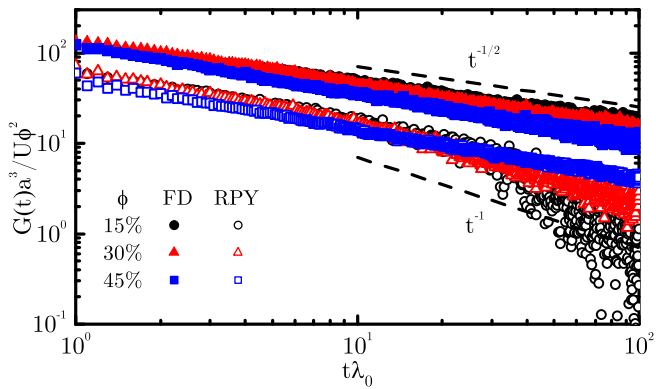}
\caption{The time-dependent shear modulus as a function of lag time after an instantaneous strain increase of $\gamma_0=1\%$ as measured for the three different volume fractions for both hydrodynamic models. Measurments were averaged over 500 realizations for each data set to reduce thermal noise. We estimate the single particle relaxation $\lambda_0$ based on the depletion well depth $U=-10k_B T$ and range of $0.1a$.\label{fig:stepstrain}}
\end{figure}

We have detailed the influence of HI on the relaxation dynamics of a colloidal gel in the linear response limit via normal mode analysis. These results are compared with data from dynamic simulations of step strain responses among particles interacting via a short-ranged attraction \cite{varga2016hydrodynamics}.  The Laplace transform of $P(\lambda)$ is the time dependent relaxation modules of the gel: $G(t)-G_e\sim\int_0^\infty P(\lambda) \exp(-t\lambda) d\lambda$, where $G_e$ denotes the infinite time value of the modulus\cite{zaccone2014linking}.  In figure \ref{fig:stepstrain} we plot the calculated time-dependent shear modulus $G(t)$, normalized by $\phi^2$ as measured in dynamic simulations using both the FD and RPY models. At short times, the stress decay reflects the fast relaxation modes due to hard core repulsion of the particles\cite{Heyes1993, varga2015hydrodynamics}. For longer lag times hydrodynamic interactions influence stress relaxation. The FD model gives $G(t)$ with a power-law scaling, $t^{-1/2}$, for $ \phi = 0.15 $. The RPY model exhibits significantly faster relaxation, as $ t^{-1}$. $G(t)$ exhibits the same asymptotic scaling predicted by the DOS for both hydrodynamic models. As the volume fraction increases, the power-laws come more into alignment, but differences between the hydrodynamic models persist as in the DOS. 

We have computed the normal modes of weak colloidal gels with different approximations for the hydrodynamic interactions between the constituent particles. Our results show that the dynamics of a gel are a sensitive function of the hydrodynamic model. Long-ranged hydrodynamic interactions enhance the rate of stress relaxation in a strained gel and play a crucial role in determining the macroscopic properties of weak colloidal gels.  With increasing volume fraction, hydrodynamic screening occurs, the majority of particles are multiply bonded, and the gel dynamics are less sensitive to the hydrodynamic model. We have confirmed the normal mode predictions by performing stress relaxation simulations of the same gels. A computational model neglecting hydrodynamic interactions will yield erroneous estimates of $G(t)$ and other related viscoelastic and mechanical properties. Future work will incorporate non-central forces between bonded particles, which could affect the modes of relaxation and dynamics of strong gels. 

The authors acknowledge helpful conversations with Professors Eric Furst and Jan Vermant, and funding provided by the ACS Petroleum Research Fund (grant no. 56719-DNI9) and the Institute for Soldier Nanotechnology at MIT.

\end{document}